# Identification of direct residue contacts in protein-protein interaction by message passing

Martin Weigt[1,2,†], Robert A. White[1,3,†], Hendrik Szurmant[3],
James A. Hoch[3,*], Terence Hwa[1,*]

[1]Center for Theoretical Biological Physics and Department of Physics,
University of California at San Diego, La Jolla, CA 92093-0374

[2]Institute for Scientific Interchange, Viale S. Severo 65, I-10133 Torino, Italy

[3]Division of Cellular Biology, Department of Molecular and Experimental Medicine,
The Scripps Research Institute, La Jolla, CA 92037

[†]These authors contributed equally to the study

*Corresponding authors: email: hwa@ucsd.edu, hoch@scripps.edu

# Abstract

Understanding the molecular determinants of specificity in protein-protein interaction is an outstanding challenge of post-genome biology. The availability of large protein databases generated from sequences of hundreds of bacterial genomes enables various statistical approaches to this problem. In this context co-variance based methods have been used to identify correlation between amino acid positions in interacting proteins. However, these methods have an important shortcoming, in that they cannot distinguish between directly and indirectly correlated residues. We developed a novel method that combines co-variance analysis with global inference analysis, adopted from use in statistical physics. Applied to a set of over 2500 representatives of the bacterial two-component signal transduction system, the combination of covariance with global inference successfully and robustly identified residue pairs that are proximal in space without resorting to *ad hoc* tuning parameters, both for hetero-interactions between sensor kinase (SK) and response regulator (RR) proteins and for homo-interactions between RR proteins. The spectacular success of this approach illustrates the effectiveness of the global inference approach in identifying direct interaction based on sequence information alone. We expect this method to be applicable soon to interaction surfaces between proteins present in only one copy per genome as the number of sequenced genomes continues to expand. Use of this method could significantly increase the potential targets for therapeutic intervention, shed light on the mechanism of protein-protein interaction, and establish the foundation for the accurate prediction of interacting protein partners.

\body

# Introduction

The large majority of cellular functions are executed and controlled by interacting proteins. With up to several thousand types of proteins expressed in a typical bacterial cell at a given time, their concerted specific interactions regulate the interplay of biochemical processes that are the essence of life. Many protein interactions are *transient*, allowing proteins to mate with several partners or travel in cellular space in order to perform their functions. Understanding these transient interactions is one of the outstanding challenges of systems biology (reviewed in (1)). The characterization of the molecular details of the interface formed between known interacting proteins is a requirement for understanding the molecular determinants of protein-protein interaction, the knowledge of which may be important for a variety of applications including synthetic biology, e.g., designing new specific interaction between proteins (reviewed in (2)), and pharmaceutics, e.g., protein interaction surfaces as novel drug targets (reviewed in (3)).

Experimental approaches to identify surfaces of interaction between proteins such as surface scanning mutagenesis and co-crystal structure generation are arduous and/or serendipitous. Co-crystal structures provide the best molecular resolution but are particular challenging to obtain for transient interaction partners. In addition, independent evidence is required to assure that the structure reflects an accurate picture of the physiologically relevant interaction.

Given the challenges of these experimental approaches, it is clear that the comprehensive identification of interaction surfaces between a large number of cellular proteins may be significantly expedited by novel computational methods. Rapid increase in the number of sequenced bacterial genomes in the past decade (resulting in more than 700 completed genome projects to date (4)) has fueled the increasing use of covariance based methods of sequence analysis for protein structure studies: Early on, these methods were largely applied to single proteins (5-9), e.g., in attempts to provide insight into tertiary structure. More recently, applications have also been made to identify interacting residues between proteins (10-13). Co-variance methods rely on the premise that amino acid substitution patterns between interacting residues are *constrained* and hence correlated. To maintain protein function, the acceptance of a deleterious substitution at one position must be compensated for by substitution(s) in the residue(s) interacting with it (14). Traditional covariance methods identify interacting residue position pairs as those exhibiting correlated substitution patterns. Applying this idea to protein-protein interaction for which the structures of the individual protein partners are known (11), one would simply look for correlation in substitution patterns between residues of the interacting partners, and identify the surfaces defined by the co-varying residues as the interacting surfaces.

However, the covariance approach has a number of shortcomings, which may significantly affect its predictive power (15). One important problem stems from the fact that correlation in amino acid substitution may arise from *direct* as well as *indirect* interactions. For example, a substitution in one position of a protein may cause conformational changes of other residues in the same protein. Such a substitution may influence the interaction between two proteins without being directly at the interface and can even occur without being proximal to the interaction surface residue at all. A classic example of this type is the allosteric effect; but indirect correlations do not require large conformational changes and may result also from cumulative effects arising from a web of small direct interactions (see below).

Traditional co-variance methods are unable to distinguish between direct and indirect correlation. A major focus of the present work is to develop a method to disentangle these correlations. Our approach is based on two premises: (i) the direct interactions are contained in the pairs of correlated residues as identified, e.g., by the covariance method, and (ii) all detected correlations in substitutions are generated by the set of direct interactions. One strategy to identifying the set of directly interacting residue positions would be to try out all possible subsets of correlated residue pairs as direct interactions. A formidable technical challenge with this approach is to work out the expected statistical correlation generated by a given set of trial direct interactions, since this itself is a very difficult *global* optimization problem (as exemplified by the notorious

"spin-glass" problem (16)). This challenge is dealt with here by applying a message-passing approach (17, 18). In recent years, insights from spin-glass physics have led to the development of generalized message-passing techniques, which have been applied successfully to a number of hard combinatorial problems such as K-SAT (19-21).

A further problem for inference is the sparsity of the information to be retrieved. The interaction surfaces are comprised of only a small subset of residues, each one being in contact with only a few surface residues of the interaction partner, and only a fraction of the interacting pairs exhibit covariance. Fortunately, for the many cases where the monomer structures of the interacting domains are already known, reliable information on even a limited number of interacting position pairs can already reveal the mode of interaction. The computational challenge is therefore to extract these few pairings from the large number of inter-protein position pairs (~$10^4$ for typical protein domains) which constitute a substantial level of background noise.

In the absence of structure information, detection of correlation between variable positions in interacting proteins therefore requires a large set of homologous protein sequences with *known* interaction partners. The number of sequenced bacterial genomes is soon to approach a number where such data could be extracted from protein pairs that are ubiquitously found in only a single copy per sequenced genome. At present however, analyses are still limited to proteins that are highly amplified in individual genomes.

In this study, we will apply our method to reveal direct interactions within the prototypical signal transduction system in bacteria, the two-component signaling (TCS) system, which is highly amplified (~10 per genome on average (22)) in order to regulate a flurry of adaptive responses to environmental and cellular cues; see (23) for a recent review. Signal detection is achieved by the sensor histidine kinase (SK) and the cellular response is mediated by the response regulator (RR), which most commonly is a transcription factor (24). The signal between the two proteins is passed via the transfer of a phosphoryl group, from a histidine residue located on the so-called HisKA domain of the SK to an aspartate residue on the RR (25). The SK and RR proteins are believed to interact specifically in most cases, and the coupled pairs are often revealed by adjacency in chromosomal location (reviewed in (26)). Over 2500 such coupled SK-RR pairs have been identified from ~300 sequenced bacterial genomes, making this system ideally suited for statistical analysis (11, 12). In addition, a large base of existing genetic and structural information — including numerous RR (e.g. (27)), two HisKA (e.g. (28)) and an exemplary co-crystal structure (29) — allows for critical evaluation of the results of statistical sequence analysis.

We present here a detailed two-stage analysis on TCS proteins. The covariance method is first used to identify the correlated residues, followed by a statistical message passing approach to infer direct coupling between pairs of residue positions. Our method distinguishes interacting residues from non-interacting ones for both SK/RR hetero-dimer and RR/RR homo-dimer interactions, with vastly improved accuracy from MI-based method without using any *ad-hoc* tuning parameters. We propose that this method will be applicable for general protein interface determination given a sufficient number of homologous protein sequences, a requirement that should soon be met by proteins present at a single copy per genome.

## Results and Discussion

**Detection of constrained positions in interacting proteins**

A multiple sequence alignment of $M = 2546$ homologue pairs was constructed for the HisKA domain of the SK and its partner RR domain by aligning with the respective hidden Markov models (see Methods). In the resulting database, chromosomally adjacent SK and RR sequences are concatenated to single sequences $\overline{A}^a = (A_1^a, A_2^a, ..., A_{N_{SK}+N_{RR}}^a)$ for $a = 1, ..., M$ such that positions $i = 1, ..., N_{SK}$ with $N_{SK} = 88$ correspond to the HisKA domain of SK, and $i = N_{SK}+1, ..., N_{SK}+N_{RR}$ with $N_{RR} = 124$ to the RR domain. Alignment gaps are included as a separate letter, so entries $A$ may assume 21 different values. Frequency counts $f_{ij}(A_i, A_j)$ are introduced for the joint appearance of amino acids $A_i$, $A_j$ in each intra- and inter-domain pair of positions $i$ and $j$, and $f_i(A_i)$ for each single position $i$.

To identify correlated positions between the SK and RR proteins of the TCS system, every position in the SK was compared with every position in the RR, and their mutual information (MI) was evaluated. This raw MI was corrected for finite-sample size effects by subtracting the average MI in a null model; see (11) and Supplementary Text. The resulting MI between positions $i$ and $j$, $MI_{ij} = MI_{ij}^{(raw)} - MI_{ij}^{(0)}$, allows for comparison of different pairs with respect to the statistical correlation of their amino-acid occupancies. Unconstrained position pairs are expected to have values close to zero.

The value of the score introduces a ranking of all pairs of positions between the two proteins. The histogram of scores allows for the self-consistent introduction of a threshold of $MI^{(t)}$ separating relevant mutual information from an exponential background signal (Supp. Fig. S1). We find 32 correlated position pairs with $MI > MI^{(t)}$. These pairs, involving 12 SK positions and 12 RR positions (listed in Supp. Fig. S2) constitute the starting point of our analysis. As will be shown below, the main results are nevertheless insensitive to the precise value of $MI^{(t)}$ used. Also, very similar results (not shown) are obtained when using other local pair-correlation measures, as e.g. a chi-squared test if residue frequencies in interacting and non-interacting protein pairs are drawn from the same distribution (13) and a likelihood ratio of the data under a correlated and a factorized model (12). Both measures identified almost the same pairings between the positions of the SK and RR domains as the high MI pairs (Fig. S2), and comparable even if slightly shuffled sets of high-ranking position pairs.

**Disentangling direct from indirect couplings**

The statistically correlated pairs are candidates for positions in contact at the protein-protein interface. However, statistical correlation does not automatically imply strong direct interaction. Imagine that position $i$ is coupled directly to $j$, and $j$ to $k$. Then $i$ and $k$ will also show correlation, without being directly coupled. The effect may become even more pronounced if there are multiple paths of weak couplings connecting $i$ and $k$. A strong correlation may emerge without the existence of any strong direct coupling linking these positions to any other residues position.

The MI score introduced above is a *local* one, in that it considers only one residue pair at a time, and compares different pairs only at the end after scores are determined.

This approach is therefore unable to disentangle direct from indirect couplings; the same is true for other local approaches, e.g., Refs. (12) and (13). To circumvent this problem, we infer a *global statistical model* $P(A_1,...,A_{N_{SK}+N_{RR}})$ describing the joint probability of the concatenated SK and RR sequence $(A_1,...,A_{N_{SK}+N_{RR}})$. This statistical model is required to satisfy two key conditions:

(i) It has to be consistent with the statistics of the data up to the level of residue pairs, i.e. the marginal distributions of the model for one or two positions have to coincide with the frequency counts $f_i(A_i)$ and $f_{ij}(A_i,A_j)$ introduced above:

$$P_{ij}(A_i,A_j) = \sum_{\{A_k|k\neq i,j\}} P(A_1,...,A_N) \equiv f_{ij}(A_i,A_j). \quad [1]$$

This condition has to hold for *all intra- and inter-protein pairs of positions* $(i,j)$, $i,j = 1,...,N_{SK}+N_{RR}$. Note that in principle higher correlations of three or more positions can be included in a similar way. However, the size of the available data set does not allow for going beyond two-residue correlations. The 21×21 elements of $f_{ij}(A_i,A_j)$ have to be estimated from the *M*=2546 sequences in the database; frequency counts for more than two positions would be very imprecise due to insufficient sample size.

(ii) To avoid over-fitting, the model has to show as few parameters as possible to meet condition (i). Application of the maximum-entropy principle yields the simplest possible (i.e. least constraint) model satisfying these conditions (30):

$$P(A_1,...,A_N) = \frac{1}{Z}\exp\left\{-\sum_{i<j} e_{ij}(A_i,A_j) + \sum_i h_i(A_i)\right\} \quad [2]$$

Model parameters are direct couplings $e_{ij}(A_i,A_j)$ between amino acid $A_i$ in position $i$ and amino acid $A_j$ in position $j$, and local biases $h_i(A_i)$ describing the preference for amino acid $A_i$ at position $i$. Determining these parameters to meet Eq. [1] is an algorithmically hard task, and can be achieved using a two-step procedure. All technical details are explained in the Supplemental Text:

1. Given a candidate set of model parameters, single- and two-residue distributions $P_i(A_i)$ and $P_{ij}(A_i,A_j)$ are estimated from Eq. [2]. This is computationally expensive, the exact summation over all possible protein sequences would require $O(21^{N-2}N^2)$ steps. Approximations can be achieved by MCMC sampling – which is expected to be very slow for 21-state variables – or more efficiently by a semi-heuristic message-passing approach (31). We use the latter approach; it reduces the computational complexity to $O(21^2N^4)$.

2. Once all $P_{ij}(A_i,A_j)$ are estimated, we can use gradient descent to adjust the coupling strengths $e_{ij}(A_i,A_j)$ (the $h_i(A_i)$ can be treated in a more efficient way explained in the Supplementary Text):

$$e_{ij}^{(new)}(A_i,A_j) = e_{ij}^{(old)}(A_i,A_j) + \Delta[f_{ij}(A_i,A_j) - P_{ij}(A_i,A_j)] \qquad [3]$$

This equation can be derived variationally within a Bayesian approach, it maximizes the joint probability of the data under Model [2], cf. Supp. Text. Since this probability is convex, it is guaranteed to converge to a single global maximum.

These two steps are iterated until Eq. [1] is satisfied within user-given precision. In the inferred model, matrices $e_{ij}(A_i,A_j)$ describe the *direct* coupling between residue positions *i* and *j*. To compare different position pairs, we propose a *scalar* measure of the coupling strength. For technical reasons (invariance with respect to gauge symmetries of Model [2] and robustness with respect to a pseudo-count introduced to regularize inference, cf. Supp. Text) a quantity called *direct information* (DI) is introduced. It measures the part of the mutual information of a position pair, which is induced by the direct coupling. Intuitively, it can be understood as the mutual information in a two-variable model for positions *i* and *j* only, which has the correct statistics of the amino acid occupancy of single positions, and coupling $e_{ij}(A_i,A_j)$ in between. The full technical definition is given in the Supp. Text.

Due to the scaling of the algorithmic complexity, the method cannot be applied simultaneously to all 212 positions of the protein alignment. Therefore the 60 positions of the protein alignment being involved in the 140 highest MI-ranking pairs (containing the 32 candidates for contact pairs identified before) are selected. The results are shown in Fig. 1A as a scatter plot of the full mutual information MI versus its direct contribution DI for the 1770 considered position pairs *(i,j)*. We observe that low MI implies low DI (lower left quadrant of Fig. 1A), but high MI does not necessarily imply high DI. DI can thus be used to rank the 32 potential links previously identified by MI. This distinction allows us to identify two groups of position pairs:

> *Group I*: This group, including the 9 pairs in the red quadrant in Fig. 1A, has both high MI and large direct coupling DI. It connects 8 SK positions with 5 RR positions. The strong links there are expected to represent physical interactions, i.e. direct contacts in the interface of the SK/RR dimer.
>
> *Group II*: This group, including the 22 pairs in the green quadrant in Fig. 1A, is densely connected by weak direct couplings (i.e. low DI). High MI between these pairs emerge from the cooperation of a multitude of such weak links. The second group contains 4 SK and 7 RR positions. They would not be expected to be in direct contact in the dimer, but instead might have a collective influence on the functionality of the SK/RR phospho-transfer interaction.

An additional set of 8 pairings (including one of the 32 high-MI pairs) are found just below the thresholds set for MI and/or DI. This group lies in the blue zone in Fig. 1A. It is expected to contain both, direct contact pairings and a few distant pairs.

The network defined by these residue pairs is shown in Supp. Fig. S2. Note that it contains many loops, so it cannot be found by dependence-tree based inference methods (as used in (12)). Its structure is found to be robust with respect to the precise details of the algorithm: The values of DI as inferred from half of the data set almost coincide with the values as inferred from the full data set; a slightly smaller but similar degree of coincidence is found if two disjoint half-size data sets are used (Supp. Fig. S3). In particular the high-scoring DI values are well reproduced.

Even though for each species only one sequenced strain was included in the database, sampling biases due to phylogenetic relations between the sequenced species exists. To evaluate whether DI values are sensitive to this phylogentic misdistribution, a reweighting procedure for potentially oversampled regions in sequence space is introduced: For each interaction SK/RR pair $\overline{A^a}$, the number $n_a$ of sequences having more than 80% sequence identity with $\overline{A^a}$ is determined, and the contribution of $\overline{A^a}$ to the frequency counts $f_i$ and $f_{ij}$ is assigned factor $1/(n_a+1)$. Global model inference is applied to determine DI. The ranking by modified DI is found to reproduce the original ranking: In between the 10 highest-ranking position pairs, one finds 10 common pairs, in between the first 20 pairs 17, in between the first 30 ranks 25 common position pairs are found (Supp. Fig. S4). Only for reweighing with respect to 60% sequence identity did part of the information get modified (not shown). This result illustrates that sampling has only small effects on the power of the proposed inference method in predicting contact pairs in interacting protein domains.

**The interaction surface of the SK/RR phospho-transfer interaction**

The validity of the above sequence-based predictions can be tested utilizing structural representatives of the SK HisKA domain (HK853 of *Thermatoga maritima*; PDBID: 2C2A (28)) and of the RR domain (Spo0F from *Bacillus subtilis*; PDBID: 1PEY (32)) as well as the co-crystal structure of Spo0F in complex with phospho-transfer protein Spo0B, both part of the sporulation phospho-relay in *B. subtilis* (PDBID: 1F51 (29)). Phospho-transferase Spo0B is a protein evolutionary related to the SK and features strong structural similarity, but is distinct in primary sequence (see below).

The HisKA domain exists as a homodimer of two helical hairpins, which form a four-helix bundle. The conserved histidine residue (phospho-donor for the RR) lies on the $\alpha$1-helix and faces away from the homodimer core. Using HK853 numbering (Fig. 1B), this residue (yellow) is at position 260. Roughly 20 residues downstream of H260 is a hairpin turn that terminates the $\alpha$1-helix and initiates the $\alpha$2-helix that runs anti-parallel to the $\alpha$1-helix. Positions of strong directly coupled residues (*Group I*) predicted to form direct contacts are 267, 268, 271, 272, 275 on the $\alpha$1-helix and 291, 294 and 298 on the $\alpha$2-helix, indicated by the red boxes in Fig. 1B. They are all found C-terminal to the active site histidine, in the vicinity of the hairpin and are exposed to the exterior of the four-helix bundle.

The RR domain forms an $\alpha/\beta$-fold consisting of a 5-stranded $\beta$-sheet surrounded by 5 $\alpha$-helices with the catalytic aspartate (receptor in phopho-transfer) nestled on the surface of one face of the fold, at position D54 using Spo0F numbering (indicated in yellow). Group I residue positions are 14, 15, 18, 21 and 22, as indicated by the red ellipses in Fig. 1B. These are all situated on the $\alpha$1-helix and exposed to the exterior of the RR domain.

SK residue positions belonging to the high MI but low DI *Group II* are 251, 252, 257, 264 (green boxes in Fig. 1B). Residue positions 257 and 264 are on the same face of the helix as the phosphorylatable histidine residue one turn N- or C- terminal, respectively. Positions 251 and 252 represent partially buried residues located at the base of the four-helix bundle. RR residue positions belonging to *Group II* are 56, 84, 87, 90, 94, 95 and 99 (green ellipses); all but one (residue 56) are localized in or around the α4-helix.

Mapping these coupled positions to the exemplary individual structures, it becomes clear that *Group I* pairings (red lines in Fig. 1B) define a mode of spatial interaction between the α1 and α2-helices of the SK and the α1-helix of the RR, bringing close together the catalytic site residues. It is, however, impossible to spatially align also the *Group II* residue positions (green lines in Fig. 1B), consistent with the notion that these do not present direct interactions according to their DI ranking. [The high MI values of Group II pairings likely reflect a dynamic role of these residues in arranging the active sites for phosphotransfer (33).]

The precise interaction mode predicted by *Group I* coupled pairings is revealed by the Spo0B/Spo0F co-crystal structure (29), which provides a structural example to measure the distances of most coupled residues (see Supp. Fig. S5 for a structural representation as well as detailed data on residue pair distances)[1]. All co-varying residues of *Group I* that can be mapped to the Spo0B structure are located in close proximity ($\leq$ 6Å) at the interaction surface between Spo0B and Spo0F. Additionally, 5 out of 6 pairings that just miss the set thresholds for $DI_{ij}$ and/or $MI_{ij}$ depicted in the blue zone in Fig. 1A are also within 6Å of each other.

From the scatter plot of the distance between a pair of residues against their DI and MI for the 408 position pairs matched to the co-crystal structure (Fig. 2A), the DI values (red symbols) are clearly seen as anti-correlated with distance. Almost all strong direct interactions correspond to short distances. Contrary, no strong correlation is observed between the distance and the MI values (blue symbols) alone. To be more quantitative, pairings were ranked according to their DI or MI values, and specificity (defined as the fraction of pairings with a minimal distance of less than 6 Å) was displayed as a function of scoring rank percentile (Fig. 2B). Whereas MI (blue line) produces the first false positive after only one true positive and rapidly drops to specificities of 30-40%, DI (red line) amazingly maintains specificity one for the top 2.5% of the 408 scoring pairs (=10 true positives).

---

[1] Despite significant structural homology, sequence homology between the Spo0B interaction domain and the HisKA domain is poor (*E* = 0.5 for HMM match to Spo0B) and only SK residues on the α1-helix can be reliably matched to Spo0B.

It can then be concluded that the combination of covariance and message passing is capable of identifying direct surface interactions from sequence data alone and that DI is a much better indicator of proximity of residues than MI. This is particularly important in instances where no clear overrepresentation of high MI scores can be observed. In those instances the DI ranking alone can be used to infer proximity.

**The interaction surface of RR homodimers**

Many proteins perform their function in bacteria as homo-oligomers, an example being transcription factors. Identifying their interaction surface poses considerable problems beyond the ones discussed above: statistical couplings of residue positions can result both from the role of a pairing as a residue contact inside the monomer structure, and as an inter-monomer contact. It is not *a priori* clear that both mechanisms lead to comparable statistical correlations, i.e. that both of them can be simultaneously detected in analyzing large sample sets of dimer-forming proteins. Even if found to be comparable, there is no intrinsic way to distinguish intra- and inter-monomer contacts: Only the knowledge of the monomer structure allows selecting candidate pairs for the interaction surface. On the other hand, the simultaneous detection of both types of statistical coupling would aid other methods in predicting tertiary and quaternary structures.

To test what kind of pairings are detectable, the global inference approach was applied to probe RR/RR interactions. Significant experimental support for a phosphorylation-dependent dimerization that increases transcription factor-DNA affinity in the largest class of RR proteins, the OmpR/PhoB class, has previously emerged (34). For probing of couplings within the phospho-transfer domains of RR proteins, the database construction was hence limited to proteins that contain both a RR phospho-transfer domain and a DNA binding domain of the OmpR/PhoB class (see Methods).

This search and alignment procedure identified more than 2000 proteins. MI scores were calculated for the 123x124/2 possible combinations of RR/RR pairings. The distribution plot of MI scores does not result in a clear anomalous tail (not shown), unlike what was observed for the SK/RR analysis (Supp. Fig. S1). Message passing was applied to calculate DI values for all pairings of the 60 positions contained in maximal MI scores, and the results were ranked according to their DI values (see Supp. Table S1 for the top 60 entries).

To evaluate the meaning of these DI rankings, the minimal atom distances of all pairs were determined utilizing three structural examples of OmpR class RR; those of *Escherichia coli* ArcA (PDBID: 1XHE (27)), PhoP (PDBID: 2PKX (35)) and *Streptococcus pneumoniae* MicA (PDBID: 1NXW (36)). For illustration, the 15 top-ranking pairs (excluding 3 pairs that were proximal in primary sequence) were mapped onto the ArcA structure (Supp. Fig. S6). As for the SK/RR analysis, a strong correlation between DI and minimal atom distance emerged (Supp. Fig. S7A), and the majority of the 60 top-scoring pairs (Supp. Table S1) are in close proximity within the monomer structures (i.e., within 6Å). Four dimer contacts are also identified (ranks 1, 3, 26 and 40). False positives (i.e. pairings with distances over 6Å) do not emerge until rank 24 and remain sparse within the top 60 pairs[2]. Quantification of specificity was determined as for the SK/RR analysis and demonstrates again that DI impressively enhances the predictive power over MI alone (Supp. Fig. S7B).

When mapping the four dimer contacts onto the three structural examples, it becomes apparent that the interactions formed by some individual contact residues are quite diverse (Fig. 3). The cluster involving pairings[3] 86:106 and 86:108 demonstrates nicely the type of residue variation that is the foundation of the covariance-based method: As shown in Fig. 3, a salt bridge is formed between E86 and R108, and a hydrogen bond connects E86 and S106 in the MicA structure. In PhoP, an aromatic stacking interaction between a tryptophan and a histidine residue (W86 and H106) can be observed. The ArcA dimer is stabilized by a hydrogen bond between E86 and N106; an additional interaction (salt bridge) is predicted between E86 and R108[4]. Similar variations exist in the pairing of residues 94 and 115, while the pairing between residues 89 and 109 is always a salt bridge between E and K. Of course, the appearance of the pair 89:109 in the high DI list shows that E:K is not the conserved pairing between these positions among all RRs, but only one of the popular residue pairings at these positions (see Supp Fig. S8). Information derived from such analysis may be exploited to design synthetic RR molecules with various degrees of cross talk with the endogenous system.

---

[2] Interestingly all false positives within the first 60 pairs include residues localized to the $\alpha$-1 helix. A rational for the occurrence of these apparent false positives is given in the caption to Supp. Table S1.
[3] ArcA numbering used throughout for clarity; for accurate MicA and PhoP numbering, deduct 2 from the ArcA numbering.

In summary, inter and intra monomer contacts lead to comparable statistical correlations. DI calculations provide constraints that could aid *de novo* structure prediction when applied to a single protein, or aid the verification or prediction of quaternary structure in cases where the monomer structure is available.

## Concluding Remarks

A novel computational method was introduced to infer structural details of protein-protein interactions based on primary sequence information. The method takes correlated residues from the covariance analysis as a starting point, and distangles correlations arising from direct vs. indirect interactions using a global inference approach implemented by a message-passing algorithm. The combination of covariance analysis and global inference impressively enhances the specificity of contact pair prediction as compared to more traditional, purely local covariance-based approaches (e.g. MI). Currently, the applicability of the method relies on the existence of ~10 structurally homologous protein sequences contained in a typical bacterial genome, due to the still limited number of sequenced genomes. With rapidly expanding genomic databases — including genes obtained via shotgun sequencing of environmental samples by the emerging field of metagenomics (reviewed in (37)) — the sample number should soon not be a limiting factor for the large majority of proteins that exist in a single copy in a genome, as long as they are widely occurring across the bacterial species.

---

[4] The possible E86-R108 salt bridge is not realized in the ArcA structure due to a likely crystallographic artifact. In the crystal lattice, residue R108 forms a salt bridge with an aspartyl residue in a neighboring ArcA dimer, a contact not available in solution (not shown).

The molecular details of the protein-protein interaction revealed may yield a large number of potential targets for antibiotic drug design in the absence of precise structural information. More broadly, the method of disentangling direct and indirect interactions presented here may also be valuable in aiding the interpretation of correlations observed in other large biological data sets, including mRNA and protein profiles, and neuronal spike activities.

## Material and Methods

### Database construction

Domains were aligned and culled from the non-redundant refseq database (release 19) (38) using HMMER (39). Only genomic data from unique species were included to avoid over-sampling of organisms with multiple sequenced strains in the database. Two overlapping sets of interacting domains, as defined by hidden Markov models (HMMs) in the Pfam database (40), were used in this study. For the SK/RR interaction study the accession numbers for the respective HMMs were PF00512 (SK) and PF00072 (RR). Functional association of these domains is inferred from chromosomal adjacency determined by GI numbers that differ by 1. For the RR/RR interaction the set of all proteins containing a PF00072 domain was restricted by requiring that the proteins also posses a specific DNA binding domain homologous to the OmpR-C domain (accession number PF00486). All HMM searches used are detecting complete domains (in contrast to a search for fragments of domains).

### Mutual information (MI) calculation

MI was measured as previously described (11); see Supp. Text for a review.

**Message passing**

The computationally hard task in the suggested global inference is the estimation of marginal distributions for single positions and pairs of positions in the sequence alignment. We used the computationally efficient but semi-heuristic message passing approach (17, 18). This approach would be exact on tree-like graphs of couplings between positions, but it is known to work efficiently also for loopy graphs. The standard formulation of message passing in terms of *belief propagation* (17, 18) estimates only single-variable marginal distributions $P_i(A_i)$. For estimating also two-variable distributions $P_{ij}(A_i,A_j)$ a recently proposed extension called *susceptibility propagation* (31) is used. Technical details are extensively exposed in the Supp. Text.

The computational cost of the approach *is $O(21^2 N^4)$* and therefore not feasible for the full amino-acid sequences having *N = 212* positions. Using a cutoff in MI, a subset of up to ca. 60 positions involved in high MI values is selected, and all pairs of these selected positions (intra- and inter-protein pairs) are considered. The inference of the parameters of the reduced model requires about 4 days of computational time on a single CPU of Dell dual quad-core 2.33GHz Xeon processor. A selection of 100 residues would require more than one month of computation. However, smaller residue sets (*N=32,40,50*) demonstrate that the qualitative results given in Results and Discussion do not depend on the choice of the MI cutoff, as soon as all nodes contained in the network of coupled residues of Fig. 3 are included (not shown).

**Direct information**

In the inferred statistical model, the direct information

$$DI_{ij} = \sum_{A_i,A_j} P_{ij}^{(dir)}(A_i,A_j) \ln \frac{P_{ij}^{(dir)}(A_i,A_j)}{f_i(A_i) f_j(A_j)} \qquad [4]$$

is calculated using the contribution $P_{ij}^{(dir)}(A_i,A_j)$ of the direct coupling $e_{ij}(A_i,A_j)$ between sequence positions *i* and *j* to the two-residue distribution. This contribution can be calculated in a hypothetical system containing only the two positions *i* and *j*: They are coupled by $e_{ij}(A_i,A_j)$ and have the correct single-variable marginals $f_i(A_i)$ and $f_j(A_j)$. DI measures the direct coupling strength between *i* and *j*, see the Supp. Text for a mathematical definition of $P_{ij}^{(dir)}(A_i, A_j)$.

## Acknowledgements

We wish to thank J. Cavanagh, U. Gerland, M. Mezard, A. Pagnani, E. van Nimwegen and I. Zhulin for discussion, and A. Beath for careful reading of the manuscript. MW and RAW are grateful to the hospitality of the NSF sponsored Center for Theoretical Biological Physics (Grants No. PHY-0822283) at UCSD where this research was initiated. This work was supported in part by the NIH through grant GM019416 to JAH and by a grant from the NAS/Keck futures initiative to TH. RAW additionally acknowledges the support of an NSF postdoctoral fellowship (DBI-0532925).

# Figure Legends

**Figure 1. The combined co-variance/message passing approach detects two groups of correlated pairs.** (A) Scatter plot of direct mutual information (DI) versus total mutual information (MI) reveals two classes of co-varying residue pairs, those with strong direct correlation found in the upper red quadrant (*Group I*) and those with low direct correlation found in the lower green quadrant (*Group II*). A group of pairings just around the border of the *MI* and/or the *DI* cutoff is highlighted in blue. (B) Direct and indirect interaction pairs depicted on exemplary structures of the HisKA and RR domain. All residues that appear in the network of pairings with $MI > MI^{(t)}$ were mapped onto the structures of HK853 from *T. maritima* (HisKA domain) and Spo0F from *B. subtilis* (RR domain). Those pairings showing strong direct correlation are depicted in red and connected by a red line and those that show low direct correlation are depicted in green and connected by a green line. Green lines connecting red residues represent low direct correlation for that particular residue pairing. For orientation, the N- and C-termini and relevant structural elements are labeled. The phosphotransfer sites H260 in HK853 and D54 in Spo0F are shown in yellow.

**Figure 2. Direct Information is inversely correlated with residue distance of pairs in the Spo0B/Spo0F co-crystal structure.** (A) Minimal atom distance for all 408 pairings that could be mapped to the Spo0B/Spo0F co-crystal structure was determined in Ångström and plotted either against direct information *DI* (red symbols) or total mutual information *MI* (blue symbols). (B) Specificity vs. rank percentile for predicting contact pairs via *DI* (red curve) and *MI* (blue curve). Specificity is defined as the fraction of pairings at the given rank percentile that are within 6Å in the Spo0B/Spo0F co-crystal structure.

**Figure 3. Direct interaction between the identified dimer contact pairs.** Four dimer contact pairings (red entries in Table S1) are localized to the $\alpha 4$- and $\alpha 5$- helices and are shown on the exemplary OmpR class RR structures of ArcA, PhoP and MicA as a dashed line. Whereas contact pairing 89:109 (ArcA numbering, for PhoP and MicA numbering deduct 2) happens to represent a salt bridge in all three structural examples shown here, the pairing 94:115 and the cluster involving pairings 86:106 and 86:108 demonstrate nicely the type of residue variation that is the basis of the covariance method; see text. Detailed analysis of the covariance among the residues involved in these four pairings are given in Supp. Fig. S8.

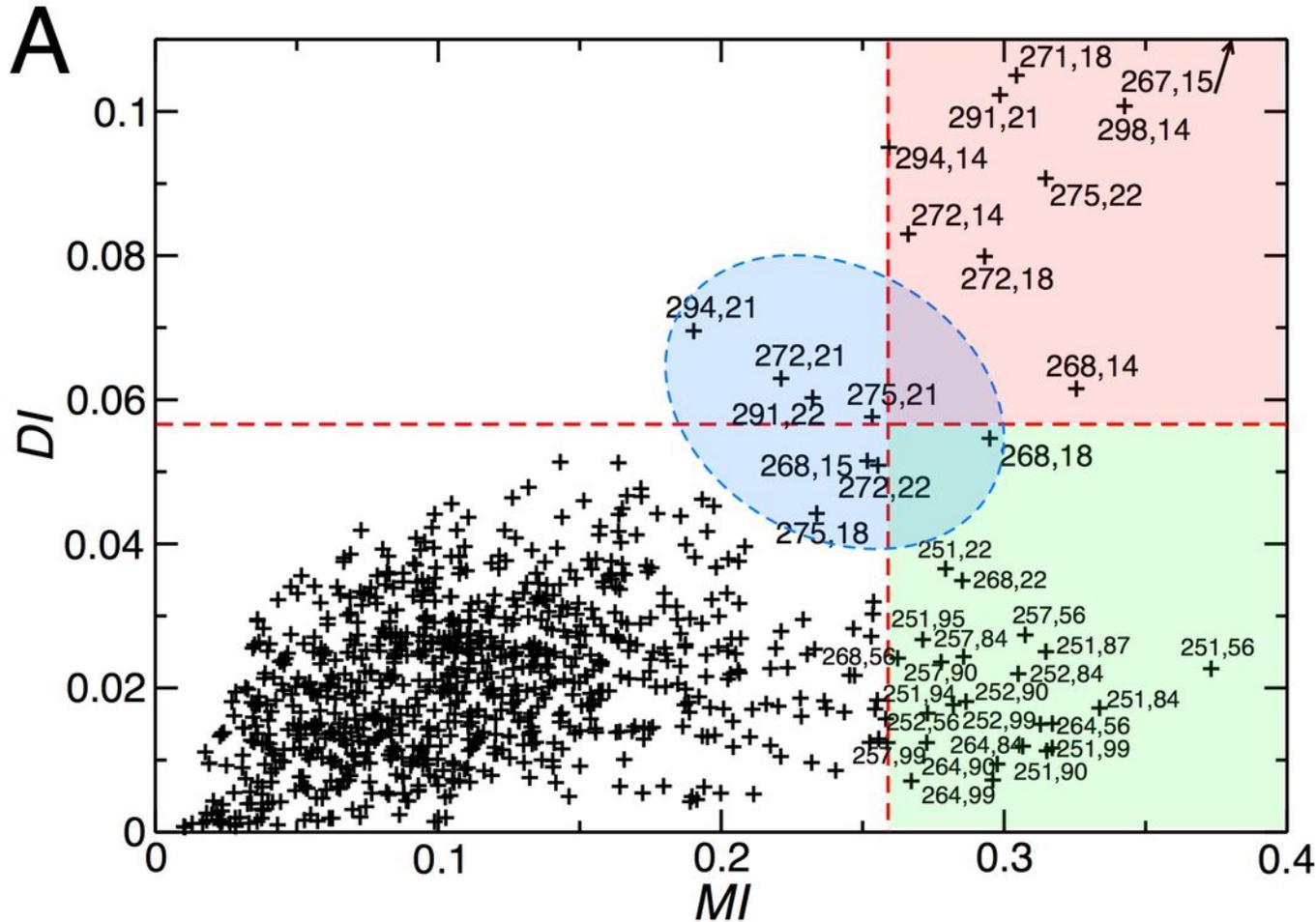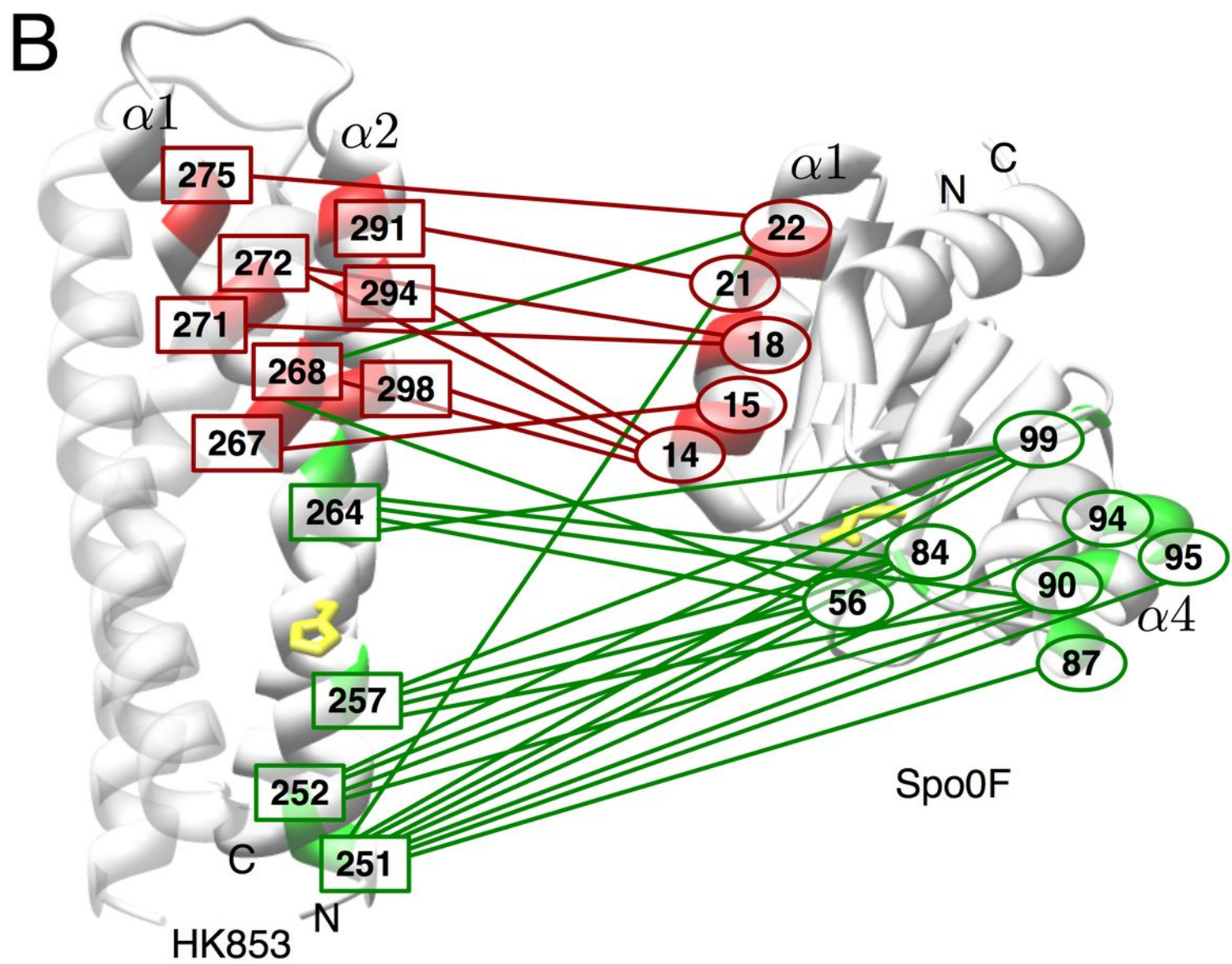

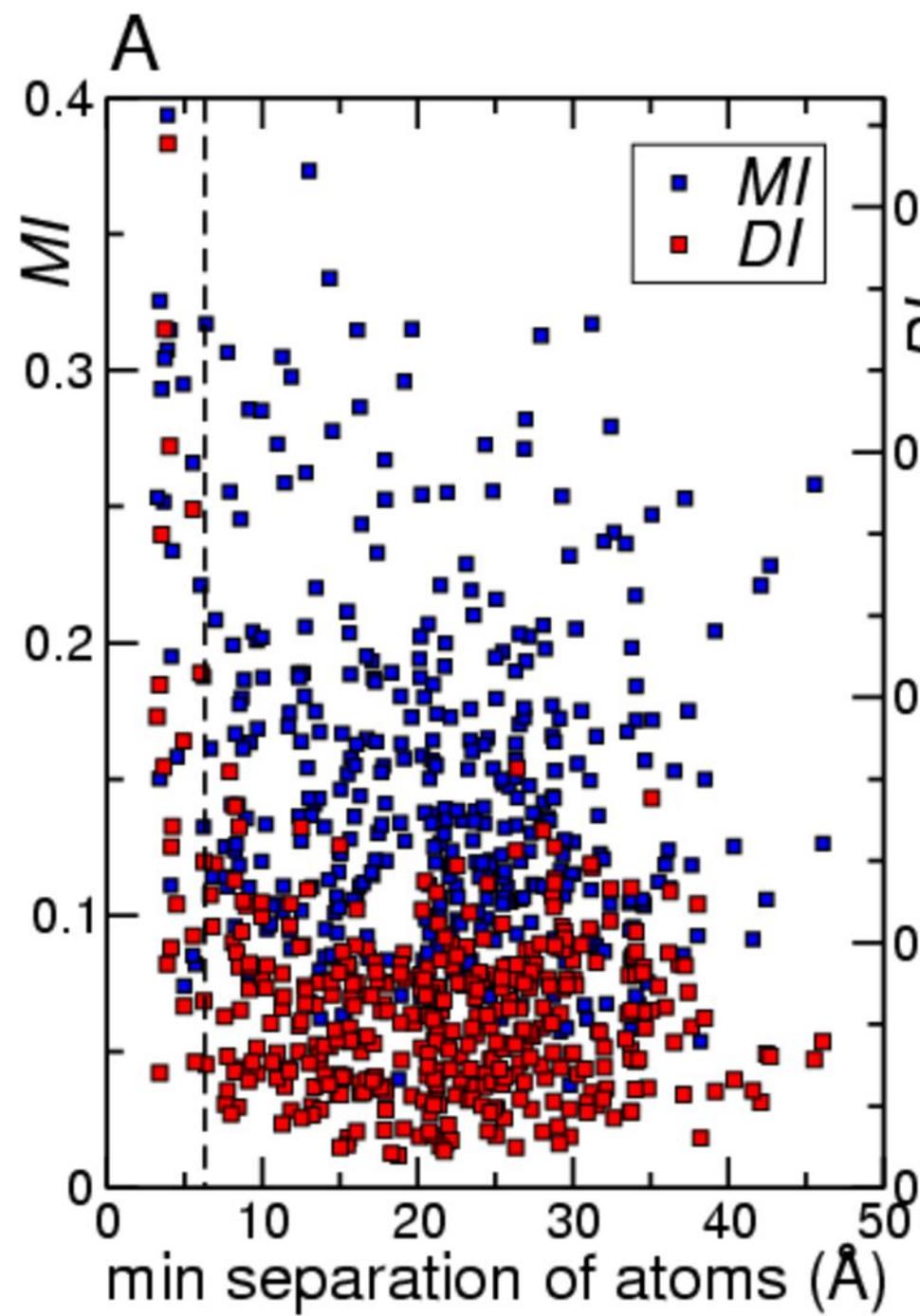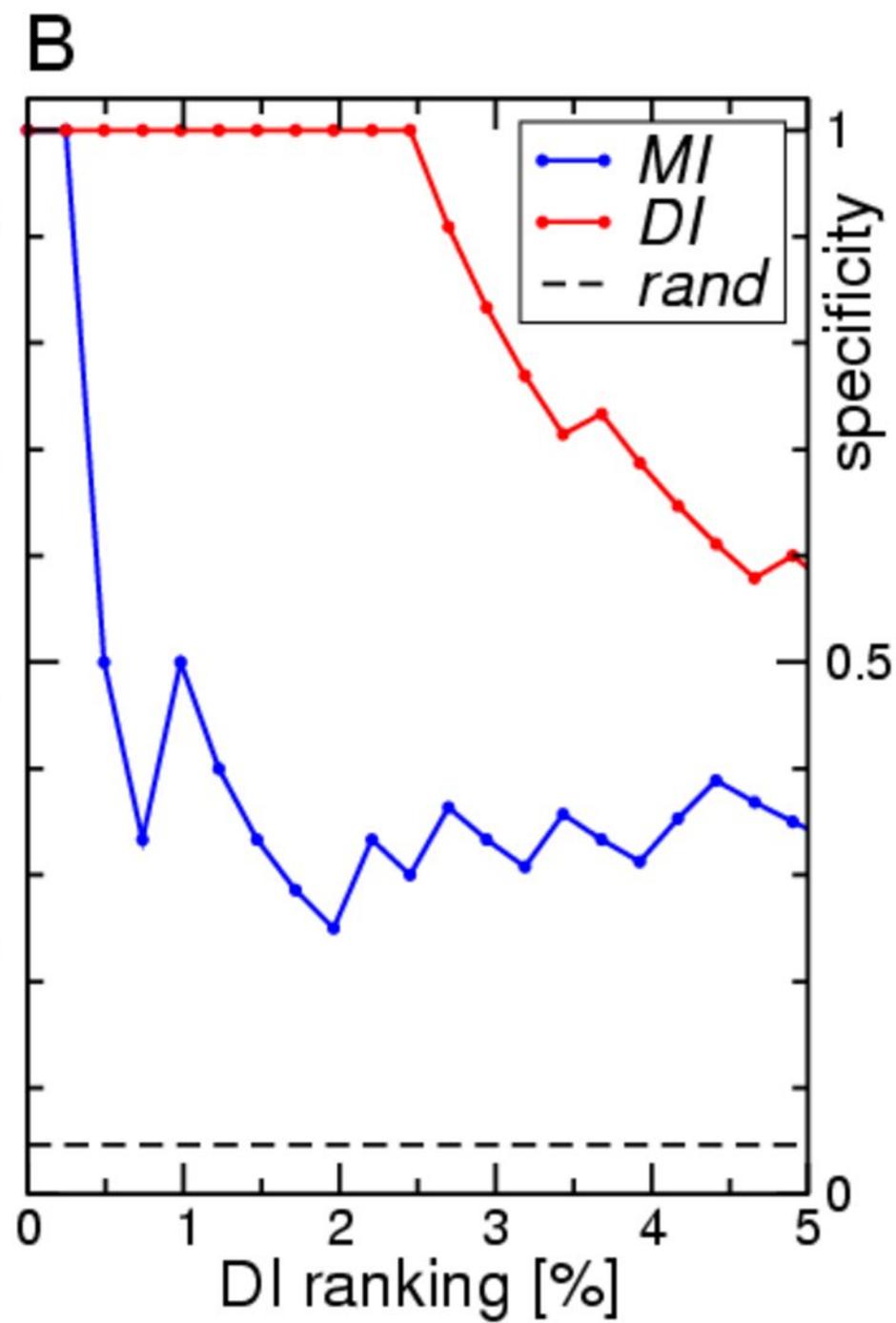

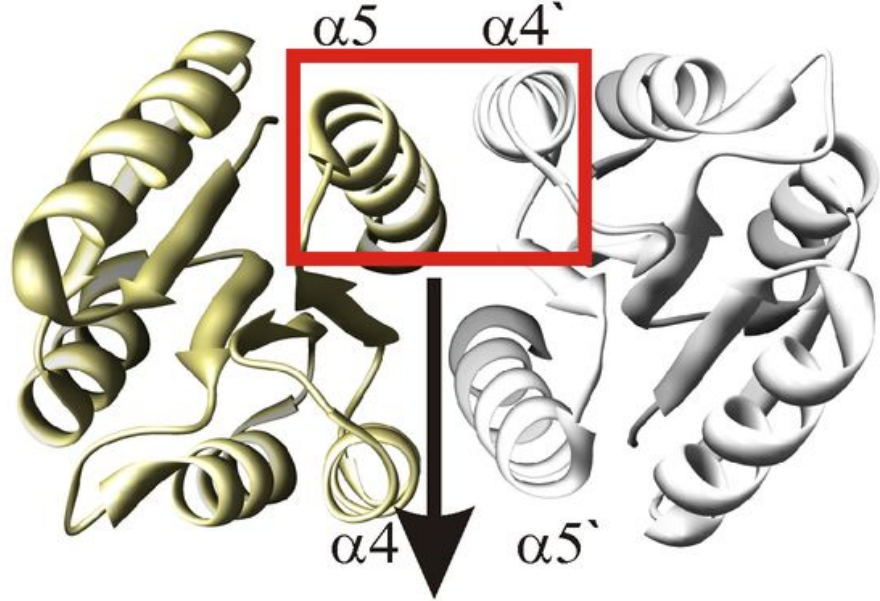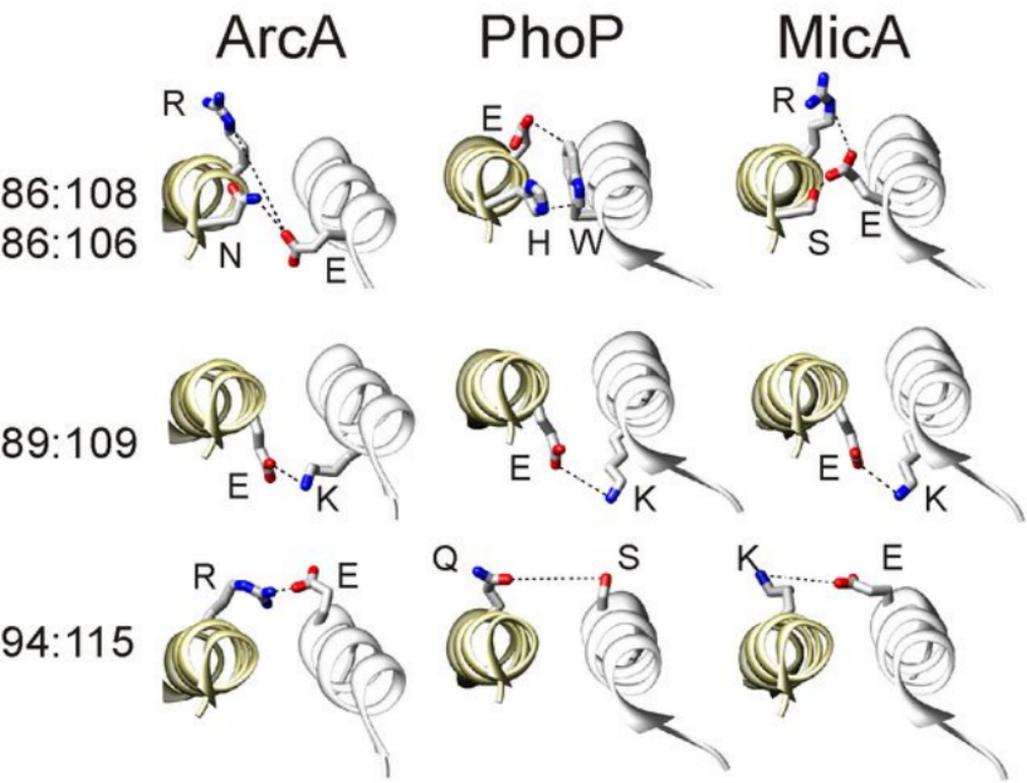